\documentclass[pra,nofootinbib]{revtex4}

\pdfoutput=1

\usepackage{color}
\newcommand{\n}{\nonumber}
\newcommand{\be}{\begin{equation}}
\newcommand{\ee}{\end{equation}}
\newcommand{\bea}{\begin{eqnarray}}
\newcommand{\eea}{\end{eqnarray}}

\begin{document}

\title{Supersymmetry  and Convection-Diffusion-Reaction equations}
\author{Choon-Lin Ho}

\affiliation
{Department of Physics, Tamkang University, Tamsui 25137, Taiwan}



\begin{abstract}

In this work we are concerned with  generating solutions of a class of Convection-Diffusion-Reaction equation from the solutions of another CDR equation through the Darboux transformations.  The method is elucidated by cases with certain types of the reaction coefficients. We have also discussed briefly supersymmetric pairs of  Convection-Diffusion-Reaction  equations connected through similarity solutions.

\end{abstract}



 \maketitle 




\section{Introduction}

The Convection-Diffusion-Reaction (CDR)  equation is an important type of second order differential equation which has found 
many important applications in physics, chemistry, astrophysics, engineering, and biology. 
 It is widely employed to model stochastic phenomena that involve the change of concentration/population of one or more substances/species distributed in space  under the influence of three processes: local reaction which modify the concentration/population, diffusion which causes the substances/species  to spread in space, and convection/drifting under the influence of external forces \cite{GK1,GK2,HM,CP,Mur}.   The well-known Fokker-Planck equation \cite{Ris,Sau} and the reaction-diffusion equation \cite{Mur} are special cases of the CDR equations. 

As with any equation in science, exact solutions of CDR equations are not easy to obtain in general.  As such, it is worthwhile to look for any method that helps find exact solutions of CDR equations.   In this work  we are concerned with  generating solutions of a class of CDR equation from the solutions of another CDR equation through the Darboux transformations.

The $(1+1)$-dimension CDR equation equation to be studied in this paper is taken to have the following general form
\be
\frac{\partial P(x,t)}{\partial t}= -\frac{\partial}{\partial x}\left(C(x,t)\,P(x,t)\right) 
+\frac{\partial}{\partial x}\left(D(x,t)\frac{\partial}{\partial x} P(x,t)\right) + R(P,x,t),
\label{CDR equation0}
\ee
where $P(x,t)$ is the particle number function,
$D(x,t), C(x,t) $  and $R(P,x,t)$ are the diffusion coefficient, the convection coefficient , and  the
reaction term, respectively.    The domains we shall consider in this paper are the real line
$x\in (-\infty,\infty)$, or  the half lines $x\in [0,\infty)$.  

In  Ref. [8] we have considered finding solutions of a FPE from the solutions of another FPE by means of the Darboux transformations \cite{Dar,Cru,MS}.  In physics literature the time-independent Darboux transformation is usually call the supersymmetric (SUSY) method \cite{SUSY}.  In Sect.\,2 we extend this approach to the CDR equations.  In Sect.\,3, we briefly discuss how SUSY pairs of CDR equations can be determined through the similarity forms of the CDR equations given in Ref. [13].

\section{CDR equations related by time-dependent Darboux transformation}

In this section, we consider the class of CDR equation equations with $D(x,t)=1$ and $R(P,x,t)= r(x,t) P(x,t)$, i.e., 
\be
\frac{\partial P(x,t)}{\partial t}=-\frac{\partial}{\partial x}\left(C(x,t)\ P(x,t)\right) +\frac{\partial^2}{\partial x^2} P(x,t) + r(x,t) P(x,t).
\label{CDR}
\ee
The diffusion coefficient $D$ is set to unity in order to make connection with supersymmetry. The real function $r(x,t)$ is the reaction coefficient. 

As we have in mind two CDR equations to be connected by the Darboux transformations, we shall denote all the defining quantities of the original CDR with a subscript ``0", and those of the partner CDR with a subscript ``1".

Let the convection coefficient of the original CDR equation be written as a space derivative of a function $W_0(x,t)$, $C_0(x,t)=-2W_0^\prime(x,t)$, where the prime indicates space-derivative.
For a solution $P_0(x,t)$ of the CDR equation, we make the ansatz
\be
P_0(x,t)=e^{-W_0(x,t)}\Psi_0(x,t).
\label{sol}
\ee
From Eq.(\ref{CDR}) one finds that the function $\Psi_0(x,t)$ satisfies a Schr\"odinger-like equation
\be
-\dot{\Psi}_0=-\Psi_0^{\prime\prime} +\left(W_0^{\prime 2} - W_0^{\prime\prime}-\dot{W}_0 - r_0\right)\,\Psi_0.
\label{Psi0}
\ee
Here the dot represents time-derivative.  It is seen that $W_0(x,t)$ determine the potential of this equation, and thus will be called the prepotential.

We shall apply the Darboux transformation to (\ref{Psi0}). Consider the Schr\"odinger equation
\be
-\dot{\Psi}_0=-\Psi_0^{\prime\prime} 
+ V_0(x,t)\Psi_0.
\label{gen}
\ee
 Suppose $\psi_0(x, t)$ is a solution of this equation. 
Then it can be checked that, for any solution $\Psi_0(x,t)$ of (\ref{gen}),  the Darboux transformed functions
 \bea
 V_1(x,t)&=&V_0-2\,(\ln\psi_0)^{\prime\prime},\n\\
\Psi_1(x,t)&=& \left(\partial_x  -(\ln\psi_0)^\prime\right)\Psi_0,
\label{Darboux}
 \eea
 also satisfy the same form of Schr\"odinger equation \cite{MS},
\be
-\dot{\Psi}_1=-\Psi_1^{\prime\prime} 
+ V_1(x,t)\Psi_1.
\label{Psi1}
\ee
The set of equations in (\ref{Darboux}) is the (time-dependent) Darboux transformation.  The function $\psi_0(x,t)$ is called the auxiliary function of the Darboux transformation.

Our strategy is as follows.  Suppose for  Eq.\,(\ref{Psi0}) we can determine the Darboux transformed equations  (\ref{Darboux}) and (\ref{Psi1}), with
\be
V_1(x,t)=W_1^{\prime 2} - W_1^{\prime\prime}-\dot{W}_1 -  r_1
\ee
for some functions $W_1$ and $r_1$, then the SUSY partner CDR is defined by the convection coefficient $C_1(x,t)=-2W_1^\prime(x,t)$ and the reaction coefficient $ r_1(x,t)$.  And for a solution $P_0(x,t)$ in (\ref{sol})  of the original CDR, a corresponding solution of the SUSY partner is
\bea
P_1(x,t)
&=& e^{-W_1}\Psi_1\n\\
&=&e^{-W_1} \left(\partial_x  -(\ln\psi_0)^\prime\right)\Psi_0 \label{P1}\\
&=&e^{-W_1} \left(\partial_x  -(\ln\psi_0)^\prime\right)\left( e^{W_0} P_0(x, t) \right).\n
\eea

Below we illustrate this construction for certain  types of the reaction coefficient $r(x,t)$.

\subsection{Case A: $r_0(x,t)=-2W^{\prime\prime}_0(x,t)$}

In this case Eq.\,(\ref{Psi0}) is
\be
-\dot{\Psi}_0=-\Psi_0^{\prime\prime} +\left(W_0^{\prime 2} + W_0^{\prime\prime}-\dot{W}_0 \right)\,\Psi_0.
\label{Psi0a}
\ee
The function $\psi_0=e^{W_0}$ is a solution of this equation, and thus can be used as an auxiliary function for the Darboux transformation giving
\bea
\Psi_1 &=&(\partial_x -W_0^\prime)\Psi_0,\n\\
-\dot{\Psi}_1 &=&-\Psi_1^{\prime\prime} +\left(W_0^{\prime 2} - W_0^{\prime\prime}-\dot{W_0} \right)\,\Psi_1.
\eea
Note the change of sign in front of the $W_0^{\prime\prime}$ term.

Now if we can find a function $W_1(x,t)$ that solves the generalized Riccati equation
\be
W_0^{\prime 2} - W_0^{\prime\prime}- {\dot W}_0
=W_1^{\prime 2} + W_1^{\prime\prime}- {\dot W}_1,
\label{R1}
\ee
then a SUSY partner CDR  equation is obtained, with $C_1(x,t)=-2W_1^\prime(x,t)$ and $r_1(x,t)=-2W_1^{\prime\prime}(x,t)$.  Given a solution $P_0=e^{-W_0}\Psi_0$ of the original  CDR equation,  the corresponding solution of the partner CDR equation is given by Eq.\,(\ref{P1}),
\be
P_1(x,t)
=e^{-W_1} (\partial_x - W_0^\prime)\left( e^{W_0} P_0(x, t) \right).
\label{P1a}
\ee

In general it is not easy to solve the generalized Riccati equation (\ref{R1}) for $W_1(x,t)$ given $W_0(x,t)$.  However, solution of (\ref{R1}) is rather easy if the prepotential $W_0(x,t)=W_0(x; a_n(t))$, where $a_n(t)$ represents collectively a set of parameters in $W_0$,  satisfies the condition of shape invariance \cite{SUSY}, namely,
\bea
&& W_0^\prime (x,; a_n(t))^2 + W_0^{\prime\prime}(x; a_n(t))\label{SI-1}\\
&=& W_0^\prime (x; a_{n+1}(t))^2 - W_0^{\prime\prime}(x; a_{n+1}(t)) + R(a_n(t)).
\n
\eea
Here $a_{n+1}(t)$ is a function of $a_n(t)$, and $R(a_n(t))$ is an $x$-independent shift function.

Shape invariance turns out to be a  sufficient condition for the exact-solvability of all the well-known one-dimensional analytically solvable quantum models, for which the prepotentials and the shift functions are time-independent.  Interestingly, for the known solvable quantum systems, the parameters $a_n$ and $a_{n+1}$ are related simply by a shift of constant. For example, $a_{n+1}=a_n + 1$ for the radial oscillator ($a_n$= angular momentum), and $a_{n+1}=a_n$ (unchanged) for the simple harmonic oscillator ($a_n$= angular frequency) \cite{SUSY}.

Using (\ref{SI-1}) we can rewrite (\ref{R1}) as
\bea
&& W_0^\prime (x,; a_n(t))^2 - W_0^{\prime\prime}(x; a_n(t)) -{\dot W}_0(x,;a_n(t))\n\\
&=& W_0^\prime (x; a_{n-1}(t))^2 + W_0^{\prime\prime}(x; a_{n-1}(t))   -{\dot W}_0(x,;a_n(t))- R(a_{n-1}(t))\label{R1a}\\
&=&  W_1^{\prime 2} (x,t)+ W_1^{\prime\prime}(x,t)- {\dot W}_1(x,t).\n
\eea
From the last equality it is seen that
a solution of (\ref{R1}) can be chosen to be 
\be
W_1 (x,t)=W_0(x; a_{n-1}(t)) + \int^t R(a_{n-1}(t))\,dt,
\ee
as long as we have
\be
{\dot W}_0(x; a_{n-1}(t))={\dot W}_0(x; a_n(t)).
\label{W0}
\ee
As pointed out in \cite{Ho},  if $a_n(t)$ and $a_{n-1}(t)$ differ only by a constant, the condition (\ref{W0}) is true if $a_n(t)$ appears in $W_0(x,t)$ only as a multiplicative factor of a function of $x$.  Of the ten solvable one-dimensional quantum models, such condition is true for: the 1d oscillator, the 3d oscillator, the Morse, the Scarf I and II, and the P\"oschl--Teller potential.

One can iterate this process to obtain a SUSY hierarchy of CDR equations defined by 
\be
W_k(x,t) = W_0(x; a_{n-k}(t))+ \int^t \sum_{s=n-k}^{n-1}\,R(a_s(t))\,dt,,~~k=1,2,\ldots
\ee
up to the lowest $a_n$ allowed by the model. 
Generalizing (\ref{P1a}), the solutions $P_k(x,t)$ of these CDR equations are related by
\bea
P_k(x, t)  =e^{-W_k }\left(\partial_x - W_{k-1}^\prime\right) \left( e^{W_{k-1}} P_{k-1}(x, t) \right),~~k=1,2,\ldots
\label{Pka}
\eea

As an example, let us take $W_0$ to be the prepotential of the 1d oscillator, 
$W_0(x,t)=\gamma(t)x^2/4$. This prepotential has the distinctive characteristics that the parameters $a_n(t)=\gamma(t)$ are all the same, and $R(a_n(t))=\gamma(t)$.  This means all the solutions $P_k(x,t)$ in the SUSY hierarchy are the solutions of the same CDR equation with $D^{(1)}=- \gamma(t) x$.

Following the discussion in \cite{Ho}, for $\gamma(t)=-1/(t+C)$, we find 
\be 
P_0(x,t)=\sqrt{\frac{t+C}{4\pi t}}\,e^{-\frac{C x^2}{4 t (t+C)}},
\ee
to be a solution of the original CDR equation.   Then from Eq.\,(\ref{Pka}), the next two solutions in the hierarchy are
\be 
P_1(x,t)\propto \frac{Cx}{t}\sqrt{\frac{t+C}{4\pi t}}\,e^{-\frac{C x^2}{4 t (t+C)}},
\ee
 and
\be 
P_2(x,t)\propto \frac{C(2Ct+2t^2-Cx^2)}{t^2}\sqrt{\frac{t+C}{4\pi t}}\,e^{-\frac{C x^2}{4 t (t+C)}}.
 \ee

\subsection{Case B: $r(x,t)=-2\dot{W}_0(x,t)$}

Eq.\,(\ref{Psi0}) in this case becomes
\be
-\dot{\Psi}_0=-\Psi_0^{\prime\prime} +\left(W_0^{\prime 2} - W_0^{\prime\prime} + \dot{W}_0 \right)\,\Psi_0.
\label{Psi0b}
\ee
Note that the last two terms in (\ref{Psi0b}) differ in the signs from those in (\ref{Psi0a}).

This time the function $\psi_0=e^{-W_0}$ is a solution of eq.\,(\ref{Psi0b}),  and the Darboux transformations based on $\psi_0$  are
\bea
\Psi_1 &=&(\partial_x +W_0^\prime)\Psi_0,\n\\
-\dot{\Psi}_1 &=&-\Psi_1^{\prime\prime} +\left(W_0^{\prime 2} + W_0^{\prime\prime}+\dot{W} \right)\,\Psi_1.
\eea

The generalized Riccati equation one needs to solve for the new prepotential 
 $W_1(x,t)$ is
 \be
W_0^{\prime 2} +W_0^{\prime\prime}+{\dot W}_0
=W_1^{\prime 2} - W_1^{\prime\prime} + {\dot W}_1,
\label{R2}
\ee
Once $W_1$ is found, the SUSY partner CDR  equation is then defined by $C_1(x,t)=-2W_1^\prime(x,t)$ and $r_1(x,t)=-2{\dot W}_1(x,t)$.  Given a solution $P_0=e^{-W_0}\Psi_0$ of the original  CDR equation,  the corresponding solution of the partner CDR equation is given by
\be
P_1(x,t)
=e^{-W_1} (\partial_x + W_0^\prime)\left( e^{W_0} P_0(x, t) \right).
\ee

As in Case A, if  the prepotentials satisfy the condition of shape invariance, then solution of (\ref{R2}) is easy. 
The generalized Riccati equation is
\bea
&& W_0^\prime (x,; a_n(t))^2 + W_0^{\prime\prime}(x; a_n(t)) + {\dot W}_0(x,;a_n(t))\n\\
&=& W_0^\prime (x; a_{n-1}(t))^2 - W_0^{\prime\prime}(x; a_{n-1}(t))   + {\dot W}_0(x,;a_n(t)) + R(a_{n}(t))\\
&=&  W_1^{\prime 2} (x,t) - W_1^{\prime\prime}(x,t) + {\dot W}_1(x,t).\n
\eea
A solution of (\ref{R2}) can be chosen to be 
\be
W_1 (x,t)=W_0(x; a_{n+1}(t)) + \int^t R(a_{n}(t))\,dt,
\ee
as long as we have
\be
{\dot W}_0(x; a_{n+1}(t))={\dot W}_0(x; a_n(t)).
\ee

Similarly, iterating this process one obtains a SUSY hierarchy of CDR equations defined by 
\be
W_k(x,t) = W_0(x; a_{n+k}(t))+ \int^t \sum_{s=n}^{n+k-1}\,R(a_s(t))\,dt,,~~k=1,2,\ldots
\ee
The solutions $P_k(x,t)$ of these CDR equations are related by
\bea
P_k(x, t)  =e^{-W_k }\left(\partial_x + W_{k-1}^\prime\right) \left( e^{W_{k-1}} P_{k-1}(x, t) \right),~~k=1,2,\ldots\n
\eea

Example of this case can be easily constructed following the description in Case A, and so will be be discussed here.

\subsection{Case C: $r(x,t)=2W_0^\prime S_0^\prime -S_0^{\prime 2} -S_0^{\prime\prime} -\dot{S}_0$}

Consider the following way to link a FPE and a CDR equation.

The FPE with a drift coeeficient $C(x,t)=-2\omega_0^\prime(x,t)$ is
\be
\frac{\partial P_{F,0}(x,t)}{\partial t}=\frac{\partial}{\partial x}\left(2\omega^\prime_0(x,t)P_{F,0}(x,t)\right) +\frac{\partial^2}{\partial x^2} P_{F,0}(x,t).
\label{FP}
\ee
Setting $P_{F,0}(x,t)=e^{S_0(x,t)}P_0(x,t)$ , where $S_0(x,t)$ and $P_0(x,t)$ are smooth functions of $x$ and $t$, we find $P_0(x,t)$ satisfies a CDR equation
\be
\frac{\partial P_0(x,t)}{\partial t}=\frac{\partial}{\partial x}\left(2W_0(x,t)^\prime\ P_0(x,t)\right) +\frac{\partial^2}{\partial x^2} P_0(x,t) +r_0(x.t)P_0(x,t),
\label{CDR1}
\ee
with
\be
W_0=\omega_0+S_0, ~~~~ r_0=2W_0^\prime S_0^\prime -S_0^{\prime 2} -S_0^{\prime\prime} -\dot{S}_0.
\label{para1}
\ee

Thus every FPE in (\ref{FP}) corresponds to a CDR equation with $W_0(x,t)$ and $r_0(x,t)$ given by (\ref{para1}) for a given $S_0$.  By choosing different functions $S_0$, one generates different CDR equations,   all corresponding to the same FPE (\ref{FP}).  This correspondence implies that, if we find a solution of the
 $P_{F,0}=e^{-\omega_0}\Psi_0$, with $\Psi_0$ satisfying \cite{Ho}
\be
-\dot{\Psi}_0=-\Psi_0^{\prime\prime} +\left(\omega_o^{\prime 2} - \omega_0^{\prime\prime}-\dot{\omega}_0\right)\,\Psi_0,
\label{Psi_FP}
\ee
then a solution of the corresponding  CDR equation (\ref{CDR1}) is $P_0=e^{-W_0}\Psi_0=e^{-S_0}P_{F,0}$.

It then follows that, if $P_{F,1}=e^{-\omega_1}\Psi_1$ is a solution of the SUSY partner FPE, where the  drift $\omega_1$ is determined by the time-dependent Darboux transformation discussed in \cite{Ho}, and $\Psi_1=(\partial_x -\omega_1^\prime)\Psi_0$, then  for a given $W_1(x,t)$, 
\bea
P_1 &=& e^{-W_1}\Psi_1\n\\
&=&e^{-W_1}\left(\partial_x -\omega_1^\prime\right) \left(e^{W_0}P_0\right)
\label{P1c}
\eea
is a solution of the corresponding SUSY partner CDR equation defined by the drift and reaction functions
\be
S_1=W_1-\omega_1, ~~~~ r_1=2W_1^\prime S_1^\prime -S_1^{\prime 2} -S_1^{\prime\prime} -\dot{S}_1.
\label{para2}
\ee
 
 As an example, let us take
 \be
\omega_0=\gamma(t) x^2/4,~~~\gamma(t)=-\frac{1}{t+C},  ~~C={\rm real\  constant.}
\ee
From Ref. [8], a solution of the FPE is
\be
P_{F,0}(x,t)=\frac{1}{\sqrt{4\pi t(t+C)}}e^{-\frac{C x^2}{4t(t+C)}}.
\ee
CDR equation corresponding to this FPE is defined by $W_0, S_0$ and $r_0$ in (\ref{para1}).  As mentioned before, there could be many possible choices of $W_0$ and $S_0$.  For illustration purpose, let us take $W_0=0$, then
\be
S_0(x,t)=x^2/[4(t+C)], ~~~ r_0(x,t)=-\frac{1}{2(t+C)}.
\ee
From $P_0=e^{-S_0}P_{F,0}$, we get a solution of this CDR equation to be
\be
P_0(x,t)=\frac{1}{\sqrt{4\pi t(t+C)}}e^{-\frac{ x^2}{4t}}.
\ee

Now we consider the SUSY partner of this CDR equation. A  choice of $\omega_1$ related to $\omega_0$ by the time-dependent Darboux transformation is $\omega_1=ax + ax^2 -\ln(t+C)/2$ ($a$ a real constant) \cite{Ho}. 
To define a SUSY partner with $\omega_1=W_1-S_1$, we take $W_1=ax$ and $S_1=\ln(t+C)/2-at^2$.  This leads to
\be
r_1(x,t)=-\dot{S}_1=-\frac{1}{2(t+C)} + a^2,
\ee
and by (\ref{P1c}), a solution of the partner CDR  equation,
\be
P_1(x,t)= \frac{x+2at}{4 \sqrt{\pi(t+C)} t^{3/2}}\,e^{-\frac{1}{4 t}(x^2+4 a xt )}.
\ee

\section{SUSY pairs  by similarity form}

Let us now discuss briefly SUSY pair of CDR equations connected through similarity solutions.

A CDR equation is said to possess scaling symmetry, if its functional form is unchanged under the scale transformation
\be
x=\epsilon^a \,\bar{x}\;\;\;,\;\;\; t=\epsilon^b \,\bar{t},~~~ \epsilon, a, b :{\rm \ real\ constants}.
\ee
  In this case,  the  functions $C, D, R$ and $P$ have the following scaling forms in terms of the similarity variable $z$:
\bea
P(x,t)&=&t^\mu y(z), ~~C(x,t)=t^\gamma\tau(z),\n\\
~~ D(x,t)&=&t^\delta \sigma(z), ~~ R(P,x,t)=t^\rho \rho(z).
\eea
 where 
  \be
z\equiv\frac{x}{t^{\alpha}}, ~~\mbox{where}~
\alpha=\frac{a}{b}\;\;\;,\; b\neq 0\; ,
\label{z}
\ee  
and $y(z), \tau(z), \sigma(z)$ and $\rho(z)$ are functions of $z$.
Scaling symmetry requires that the exponents are linked by \cite{HY}
\be
\gamma=\alpha-1,~~ \delta=2\alpha-1,~~\rho=\mu-1.
\ee
Hence $\alpha$ and $\mu$ are the only two independent scaling exponents of the CDR  equation.
With these, the CDR equation is reduced  to an ordinary differential equation \cite{HY}
\be
\sigma y''+(\sigma' +\alpha\,z-\tau )\,y' -(\tau'+\mu)\, y+\rho=0.
\label{ODE}
\ee
Here ``prime" represents derivative with respect to $z$.

Let us now consider the special case where $\sigma(z)=1, \tau(z)=\alpha z$ and $\rho(z)=-\Phi(z) y(z)$.   Eq.\,(\ref{ODE}) then reduces to
\be
-y''+(\Phi+\mu +\sigma) y=0.
\label{ODE1}
\ee
If we set  $V - E\equiv \Phi+\mu +\sigma$, where $E$ is a real constant, then (\ref{ODE1}) is recast into the form of a Schr\"odinger equation
\be
-y''+(V-E) y=0.
\label{ODE2}
\ee

It is now clear that this form permits us to derive a SUSY partner of the CDR equation by the time-independent version of the Darboux transformation (\ref{Darboux}). For any solutions $y_0(z)$ and $y(z)$ of (\ref{ODE2}), the functions
 \bea
 {\widetilde V}(z)&=&V-2\,(\ln y_0)^{\prime\prime},\n\\
{\widetilde y}(z)&=& \left(\partial_x  -(\ln y_0)^\prime\right) y,
\label{Darboux1}
 \eea
 also satisfy the same form of Schr\"odinger equation (\ref{ODE2}) with the same value of $E$.
 Here $y_0(z)$ serves as the auxiliary function of the Darboux transformation.  The SUSY partner CDR equation is then defined by the functions
 (recall $\sigma(z)=1, \tau(z)=\alpha z$)
\bea
&&{\widetilde P}(x,t)=t^\mu {\widetilde y}(z), ~~{\widetilde C}(P,x,t)=t^\gamma (\alpha z),
~~ {\widetilde D}(P,x,t)=t^\delta, \n\\
&&~~ {\widetilde R}(P,x,t)=-t^\rho\, \widetilde{\Phi}(z)\, {\widetilde y}(z),~~~ ~~\widetilde{\Phi}(z)={\widetilde V}(z)-E-\mu-\alpha.
\eea
 
It is not hard to work out examples by matching (\ref{ODE2}) with the solvable SUSY quantum systems listed in [12].


\section{Summary}

In this work  we have discussed a way to obtain solutions of a class of CDR equation from those of another CDR equation, by means of the Darboux transformation.   Cases with certain types of the reaction coefficients are presented. We have also discussed briefly SUSY pairs of CDR equation connected through similarity solutions.

We have not considered the case with $x$-independent reaction coefficient, i.e., $r(x,t)=r(t)$, as this case can be transformed into a FPE by a phase transformation.  In fact, letting
\be
P(x,t)=e^{\int^t r(t) dt}\,P_{F}(x,t),
\ee
one finds $P_{F}(x,t)$ satisfies the FPE.  Thus one can follow the discussion in Ref. [8] to determine the SUSY partners of the CDR equation.


\acknowledgments
The work is supported in part by the Ministry of Science and Technology (MoST)
of the Republic of China under Grant  MOST 110-2112-M-032-011.


\end{document}